\documentclass{PoS}

\usepackage{subfig}
\usepackage{float}

\title{LHC phenomenology of light pseudoscalars in the NMSSM}

\ShortTitle{LHC phenomenology of light pseudoscalars in the NMSSM}

\author{\speaker{Nils-Erik Bomark}\\%
        University of Agder, Kristiansand, Norway\\
        E-mail: \email{nilserik.bomark@gmail.com}}

\author{Stefano Moretti\\
        University of Southampton, Southampton, UK\\
        E-mail: \email{s.moretti@soton.ac.uk}}

\author{Shoaib Munir\\
        APCTP, Pohang, South Korea\\
        E-mail: \email{s.munir@apctp.org}}

\author{Leszek Roszkowski\\
	\thanks{On leave of absence from the University of Sheffield, UK.}
        NCBJ, Warsaw, Poland\\
        E-mail: \email{leszek.roszkowski@fuw.edu.pl}}

\abstract{After the discovery of the 125 GeV Higgs boson, the Next-to-Minimal Supersymmetric Standard Model (NMSSM) has become more interesting as a model for new physics since new tree-level contributions to the Higgs mass makes it easier to accommodate the relatively high measured value, as compared to the MSSM.\\ One very distinctive feature of the NMSSM is the possible existence of a light singlet-like pseudoscalar. As this pseudoscalar may be lighter than the discovered Higgs boson without conflict with data, it may lead to LHC signatures rather different to what is usually searched for in terms of new physics.\\ In these proceedings we will discuss studies concerning the discoverability of such light pseudoscalars.  It is demonstrated that heavier scalars decaying to pairs of pseudoscalars or pseudoscalars and $Z$ bosons may lead to discovery in a large part of the parameter space. This is especially important for the non-SM like of the two lightest scalars, as it may have an almost 100\% branching ratio for decay into pairs of pseudoscalars. In such a case the discussed channels might be our only means of discovery, also for the scalar.}

\FullConference{The European Physical Society Conference on High Energy Physics\\
		22--29 July 2015\\
		Vienna, Austria}

\begin{document}

\section{Introduction}
Despite the ever tighter exclusion limits on sparticle masses coming from the LHC, supersymmetry still remains our best bet for physics beyond the Standard Model (SM). It is worth pointing out that, although the mass of the discovered Higgs boson\cite{Aad:2012tfa,Chatrchyan:2012ufa} is a bit high for supersymmetry, it is consistent with the Minimal Supersymmetric Standard Model (MSSM).

The MSSM, though, suffers from a problem in the dimensionful supersymmetric $\mu$ term which for phenomenological reasons has to be of the same scale as the soft supersymmetry breaking terms. As these terms are a priory unrelated, this poses a problem for high scale model building.

One way to enforce the similarity in scale between the $\mu$ term and the soft supersymmetry breaking terms is to forbid the $\mu$ term and instead introduce a gauge singlet scalar superfield which then can generate the $\mu$ term by getting a Vacuum Expectation Value (VEV) at the Electroweak scale. This VEV will be generated from soft supersymmetry breaking terms and hence the similarity of scales comes out naturally.

This is the idea behind the Next to Minimal Supersymmetric Standard Model (NMSSM) \cite{Ellwanger:2009dp}. This model has enjoyed a renewed interest with the discovery of the, for the MSSM, somewhat heavy Higgs boson, since the NMSSM contains new tree level contributions to the Higgs mass and hence allows for a 125 GeV Higgs while keeping the fine-tuning at acceptable levels.

With the additional scalar superfield, the NMSSM has three neutral scalars, $H_1,H_2,H_3$ where $M_{H_1}<M_{H_2}<M_{H_3}$, and two pseudoscalars, $A_1,A_2$ with $M_{A_1}< M_{A_2}$, as well as a charged scalar $H^\pm$. The discovered SM-like Higgs, $H_{\rm SM}$, is identified as the doublet-like of $H_1$ or $H_2$.

As we are looking for light pseudoscalars, $A_1$ will always be singlet-like and its mass is basically a free parameter so we will take it to be lighter than $H_{\rm SM}$.
Such light pseudoscalars can be detected in several ways. If $M_{A_1}< 10$ GeV they might show up in meson decays. One could also consider direct production, but that seems unlikely to work; only $b\bar b A_1$ production is of interest but at least for $M_{A_1}> 10$ GeV, the dominant decay channel is $A_1\to b\bar b$ rendering a detection very challenging \cite{Bomark:2014gya}.

The remaining hope is indirect production through the decay of heavier particles. In the following we will discuss studies \cite{Bomark:2014gya,Bomark:2015fga} of the detection prospects from heavier scalars decaying to $A_1A_1$ or $A_1Z$.

\section{Scan}
In order to assess the discovery prospects of a light pseudoscalar in the NMSSM parameter space, we employ Bayesian scans using MultiNest-v2.18 \cite{Feroz:2008xx} coupled with NMSSMTools-v4.2.1~\cite{Ellwanger:2006rn},  Higgsbounds-v4.1.3 \cite{Bechtle:2008jh,Bechtle:2011sb,Bechtle:2013gu,Bechtle:2013wla}, SuperISO-v3.3 \cite{superiso} and micrOMEGAs-v2.4.5 \cite{micromegas}.

To cover all possibilities, we use two scans; one focusing on what we call the ``naturalness region'' with low $\tan\beta$ (the ratio of the vacuum expectation values of the two  Higgs doublets) and large $\lambda$ (the coupling constant for the singlet scalar to the Higgs doublets) to maximise the tree-level NMSSM specific enhancements of the SM-like Higgs mass, and one scan using wider parameter ranges. We also use separate scans for $H_1$ and $H_2$ being SM-like.



To make sure the SM-like Higgs is acceptable, we require it to be between 122 and 128 GeV where the large range is due to large theoretical uncertainties. We also require the points to comply with the Higgs signal rates for $H_{\rm SM}\to \gamma\gamma$ and $H_{\rm SM}\to ZZ$ as given by CMS in \cite{CMS-PAS-HIG-14-009}. We do not use ATLAS data here since they at the time of the scan had large deviations from SM values that have later disappeared \cite{ATLAS-CONF-2014-009,Aad:2014eha}.

\section{LHC studies}
To estimate the LHC reach in the studied channels we use MadGraph5\_aMC$@$NLO \cite{Alwall:2014hca} to generate  parton level backgrounds and then Pythia 8.180 \cite{Sjostrand:2007gs} and FastJet-v3.0.6 \cite{Cacciari:2011ma} for signal generation, hadronisation and jet clustering. To optimise the sensitivity at low pseudoscalar masses we employ the jet substructure methods of \cite{Butterworth:2008iy}.

Our studies include all channels where scalars are produced and decay to $A_1A_1$ or $A_1Z$, however, as explained in \cite{Bomark:2014gya}, $H_{1,2}\to A_1Z$ and $H_3\to A_1A_1$ have too small rates to be of any interest. This means that we focus on the lighter scalars $H_{1,2}$ decaying to $A_1A_1$ and $H_3\to A_1Z$. The latter is an interesting channel for somewhat heavier pseudoscalars that have not received much attention --- due to the small couplings between $H_3$ and the weak vector bosons, this channel can only be studied through gluon fusion (GF) production of the scalar. For the lighter scalars, also vector boson fusion (VBF) as well as Higgsstrahlung ($ZH$ and $WH$) production can be of interest, though the higher rates of GF seems to always be more important than the lower backgrounds of the other channels \cite{Bomark:2015fga}.

Since we are focusing on $M_{A_1}>10$ GeV, BR$(A_1\to b\bar b)$ is always around 0.9 making 4b-jets the dominant final state for the $A_1A_1$ channel --- however, the reduction in background in the 2b2$\tau$\footnote{Also 2b2$\mu$ could be interesting, that will be the topic of future studies.} final state more then compensates for the factor 9 in the signal rate making it the best channel if the initial scalar is produced through GF or VBF. With $WH$ and $ZH$ the backgrounds are sufficiently suppressed by the additional vector bosons (that we require to decay leptonically) so that the higher rates of $4b$ channel is more beneficial.

\section{Results}
\begin{figure}[tbp]
\centering
\subfloat[]{%
\includegraphics*[width=0.45\textwidth]{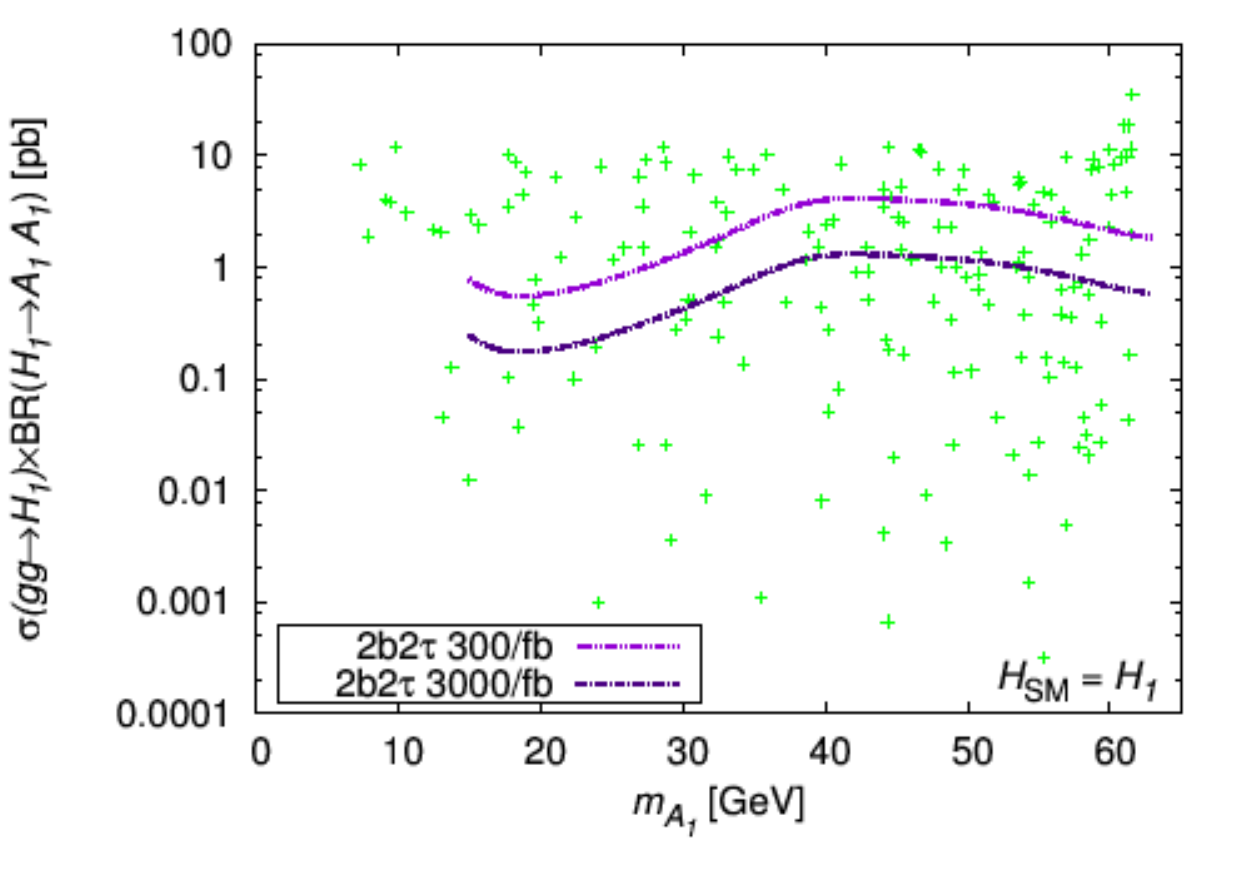}
}%
\hspace{0.5cm}%
\subfloat[]{%
\includegraphics*[width=0.45\textwidth]{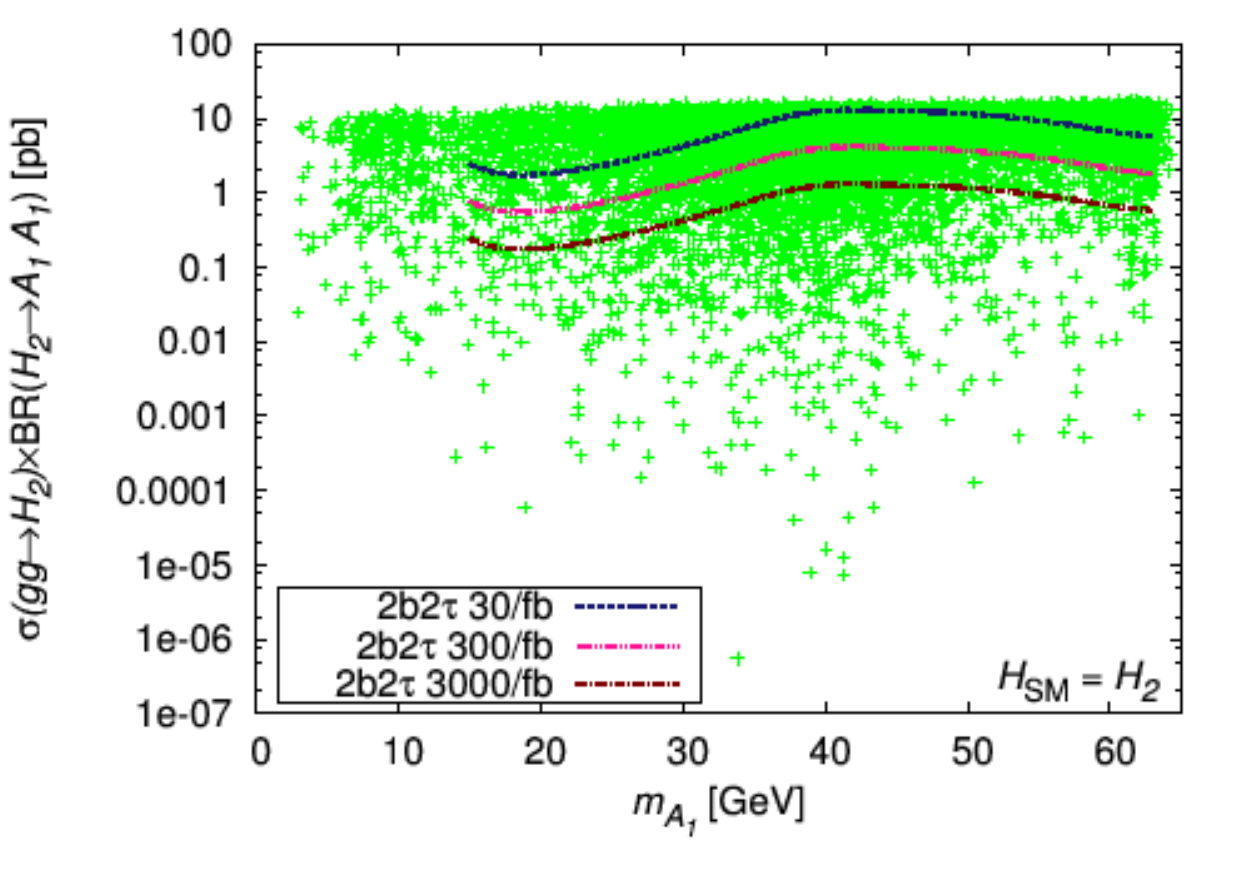}
}
\caption{LHC reach in $H_{\rm SM}\to A_1A_1$  for $H_1=H_{\rm SM}$(left) and $H_2=H_{\rm SM}$(right). The curves show 5$\sigma$ discovery reach for LHC-14.}
\label{fig:hSM}
\end{figure}

If we look at the channels with $H_{\rm SM}\to A_1A_1$, there is a problem in getting points where the channel is kinematically open, especially in the naturalness region this is an issue as $\lambda$ here is large and hence BR$(H_{\rm SM}\to A_1A_1)$ usually becomes so large that it suppresses other Higgs decays below experimentally acceptable rates, this is especially true when $H_1$ is SM-like. This can be seen in Figure~\ref{fig:hSM}, where the left panel is very scarcely populated; the right panel has better coverage but also here can we see a clear upper limit on the cross-section.

\begin{figure}[tbp]
\centering
\subfloat[]{%
\includegraphics*[width=0.45\textwidth]{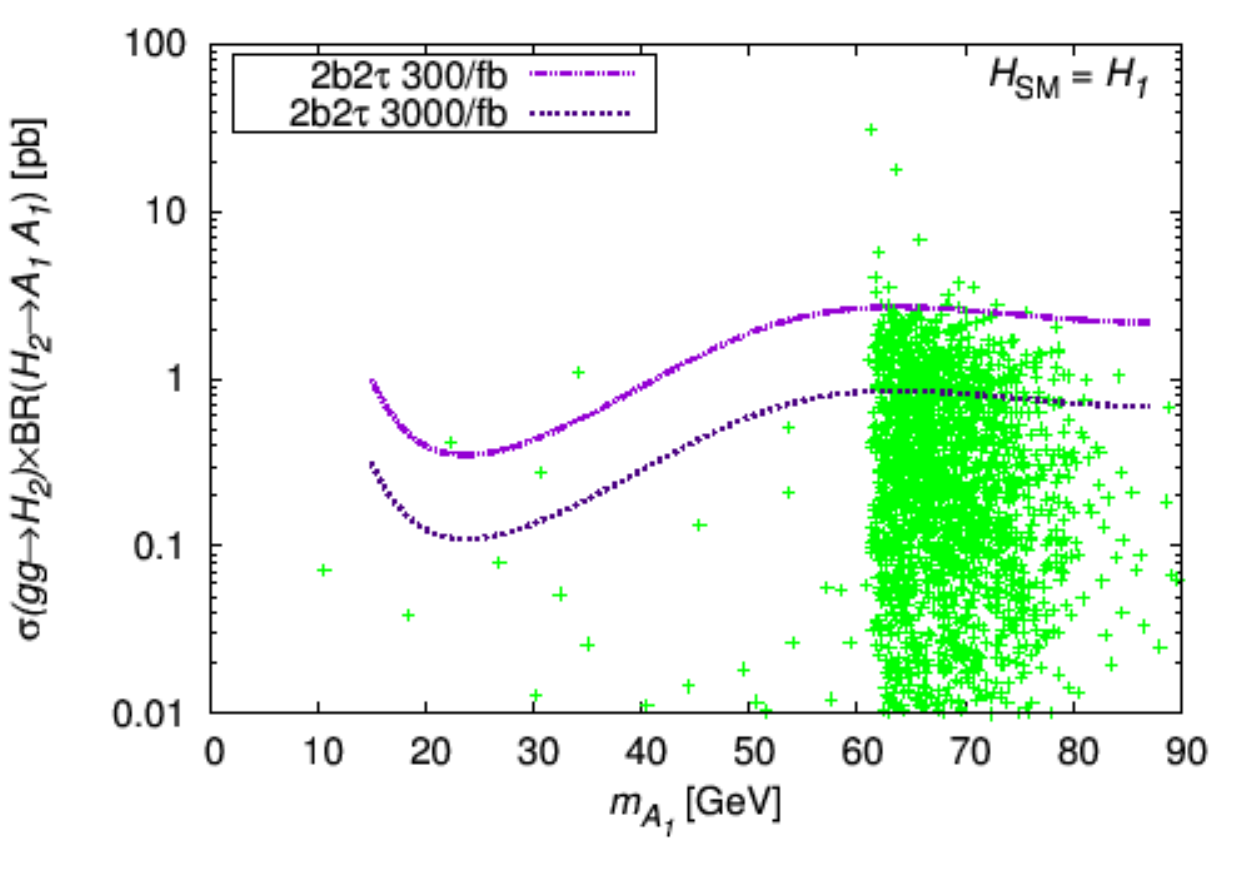}
}%
\hspace{0.5cm}%
\subfloat[]{%
\includegraphics*[width=0.45\textwidth]{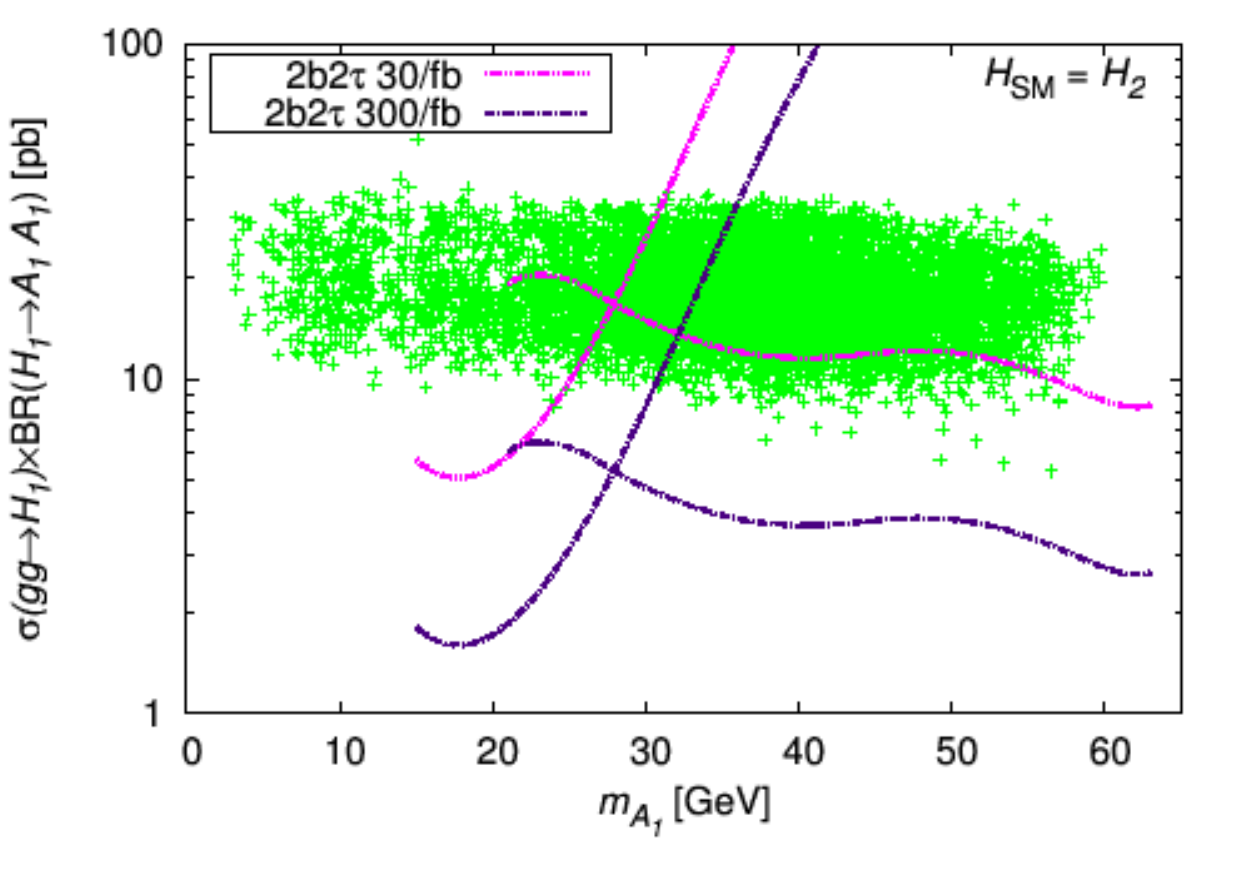}
}
\caption{LHC reach in $H_{\rm non-SM}\to A_1A_1$  for $H_1=H_{\rm SM}$(left) and $H_2=H_{\rm SM}$(right). The curves assume scalar masses of 175 (left panel) 100 (right panel low mass curves, uses jet substructure) and 125 GeV (right panel curves reaching higher masses).}
\label{fig:hnonSM}
\end{figure}

Perhaps even more interesting is the possibility to see the non-SM one of $H_1$ and $H_2$ ($H_{\rm non-SM}$) decay to $A_1A_1$. Since $H_{\rm non-SM}$ has no lower limits on other signal rates, BR$(H_{\rm non-SM}\to A_1A_1)$ may well be close to one and hence this might be our best (or only) chance of discovering also this scalar. As can be seen in Figure~\ref{fig:hnonSM} the prospects for such discovery are rather good: especially when $H_2$ is SM-like, this channel can be very promising.

\begin{figure}[tbp]
\centering
\subfloat[]{%
\includegraphics*[width=0.45\textwidth]{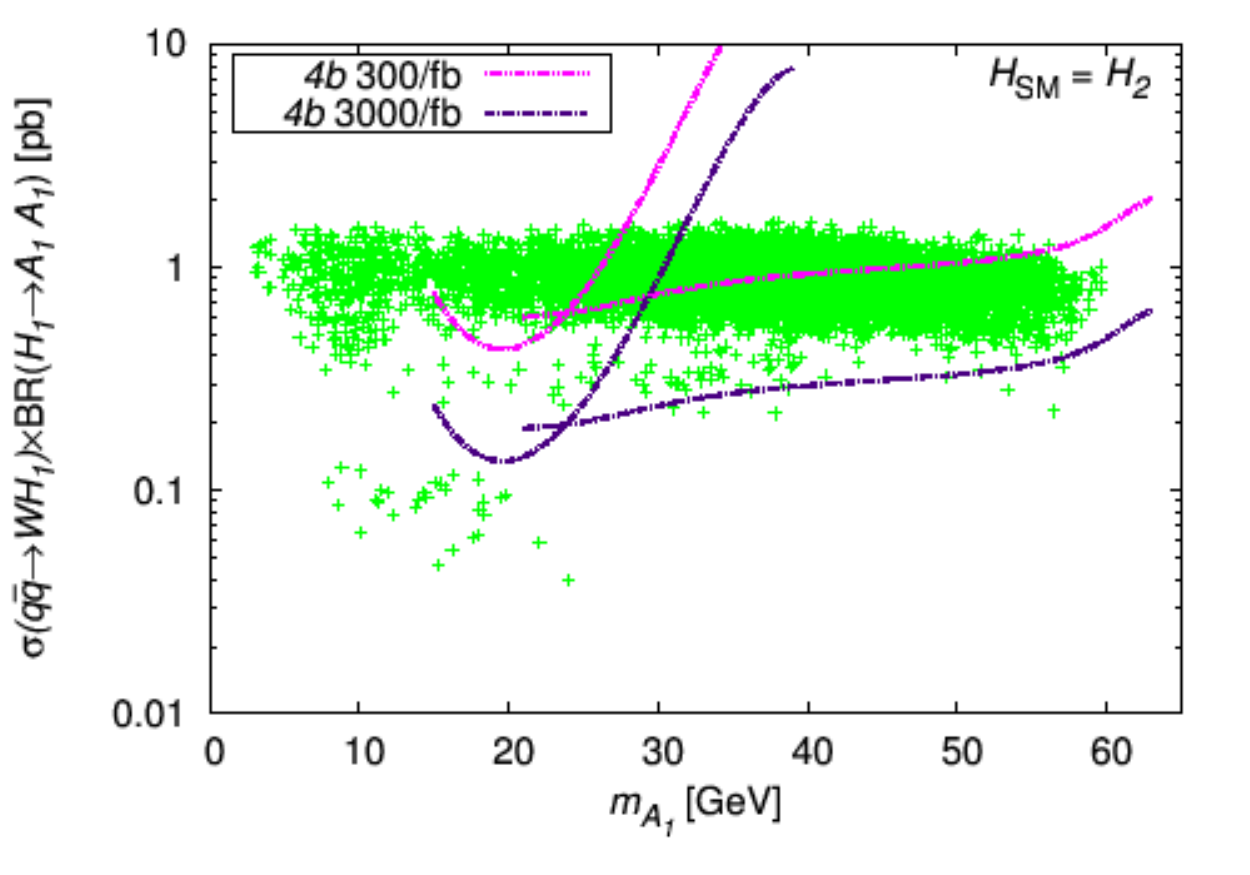}
}%
\hspace{0.5cm}%
\subfloat[]{%
\includegraphics*[width=0.45\textwidth]{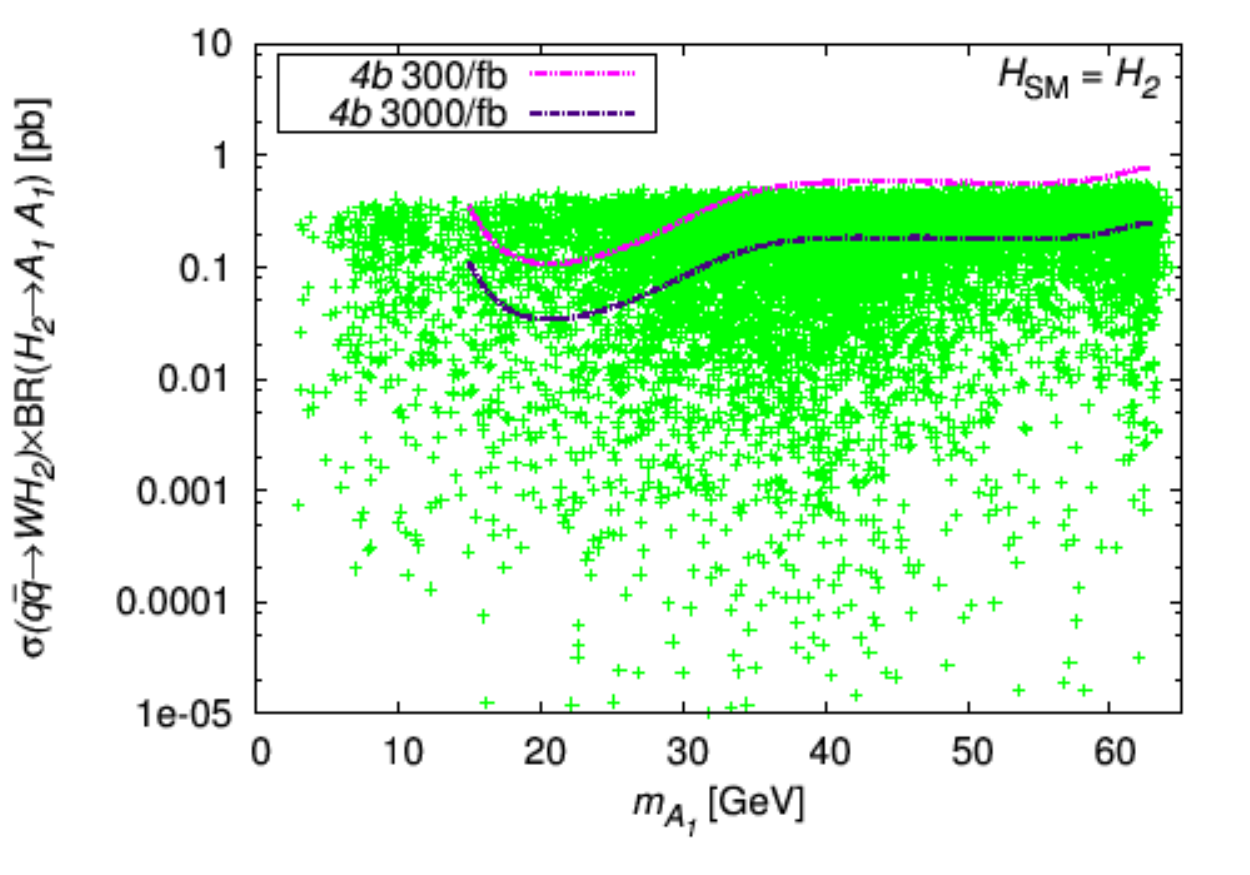}
}
\caption{LHC reach in $H_1\to A_1A_1$ (left) and $H_2\to A_1A_1$ (right) for $H_2=H_{\rm SM}$ in the $WH$ channel. The curves use the same parameters as those of figure~\protect\ref{fig:hSM} and~\protect\ref{fig:hnonSM}.}
\label{fig:WH}
\end{figure}

While GF production of the initial scalar gives the highest signal rates, it also have large backgrounds, therefore it could be interesting to compare to the reach in other channels. Of the other possibilities $WH$ seems like the most promising; $ZH$ has an almost negligible background due to the requirement of a leptonically decaying $Z$ but the signal is also very small, so $WH$ amounts to a better compromise between the two. However, as can be seen in Figure~\ref{fig:WH}, the sensitivity is clearly worse than for GF. One should remember though, that this could be an important complement, especially to measure vector boson couplings of the $H_{\rm non-SM}$.

\begin{figure}[tbp]
\centering
\subfloat[]{%
\includegraphics*[width=0.45\textwidth]{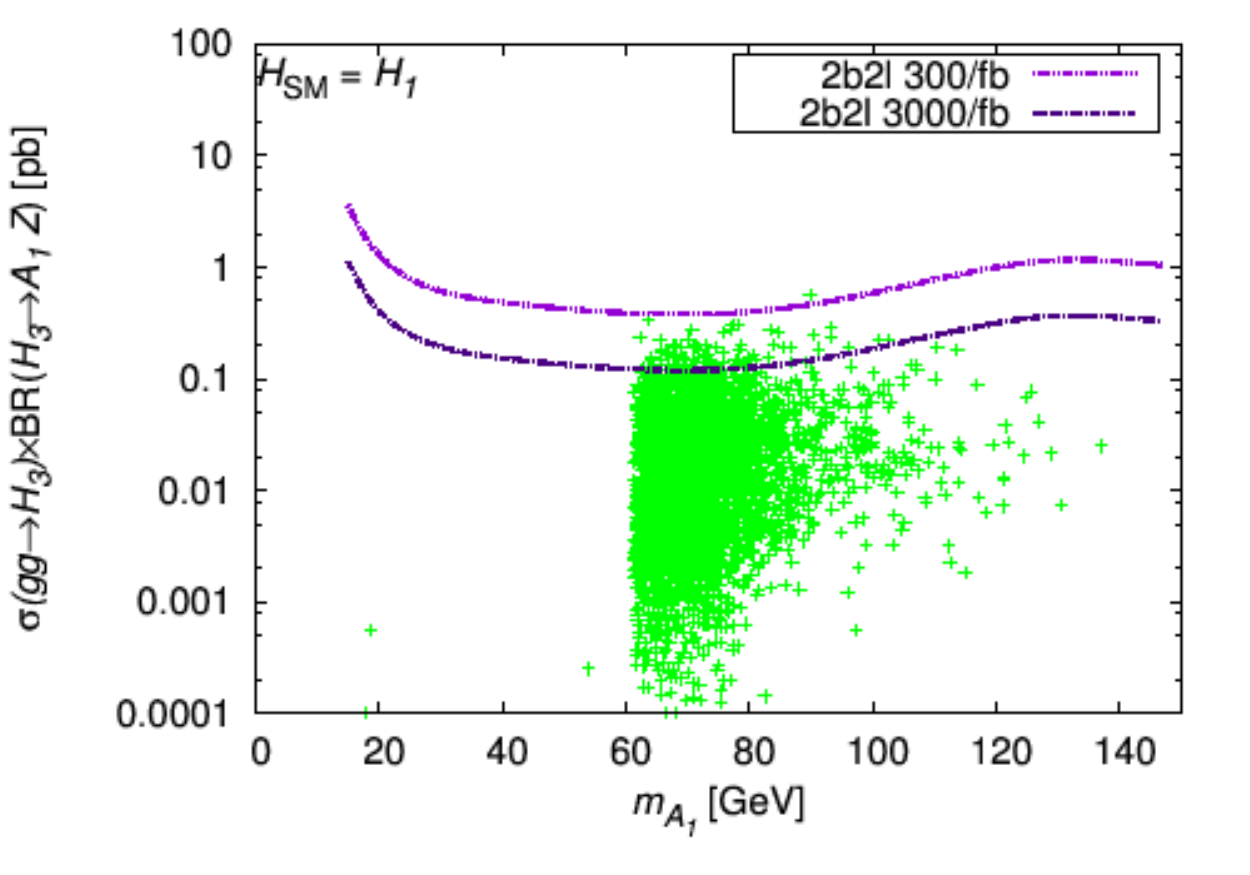}
}%
\hspace{0.5cm}%
\subfloat[]{%
\includegraphics*[width=0.45\textwidth]{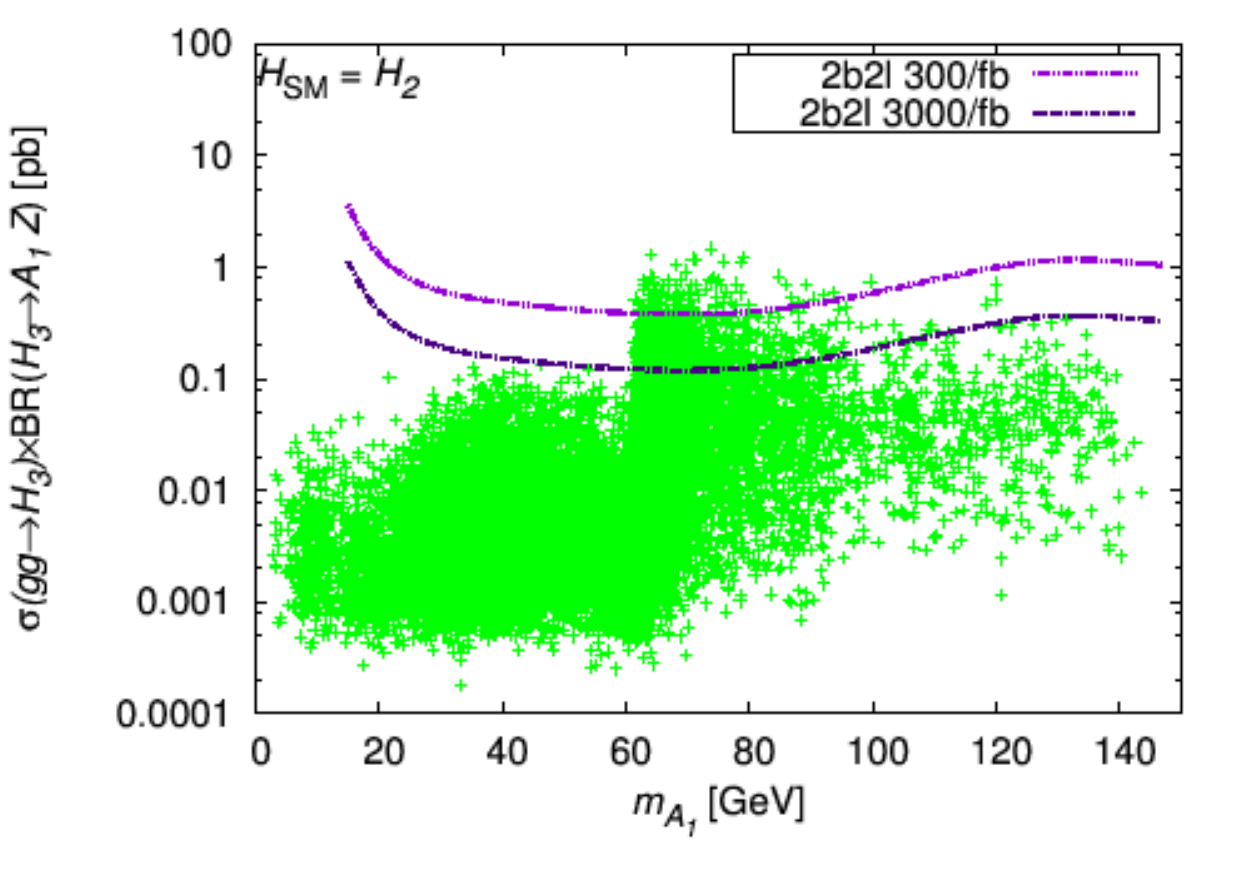}
}
\caption{LHC reach in $H_3\to A_1Z$ for  $H_1=H_{\rm SM}$(left) and $H_2=H_{\rm SM}$(right). The curves assume a scalar mass of 350 GeV.}
\label{fig:h3A1Z}
\end{figure}

For somewhat heavier $A_1$s we have to rely on the $H_3\to A_1Z$ channel, which, as can be seen in Figure~\ref{fig:h3A1Z}, does show some promise. It should be remembered that, since this work focuses on light pseudoscalars, the cuts used are generic and rather soft, so a more detailed study of the effects of harder cuts may significantly improve the scope of this channel.

The prospects in all channels are summarised in Table~\ref{tab:Channels}.

\begin{table}[tbp]
\begin{center}
  \begin{tabular}{|c|l|c|c|}\hline
  Production mode  & Channels & Accessibility & Range\,(GeV) \\\hline
  $b\bar b A_1$  &  ---  & \textcolor{red}{x} &\\
  $H_1\to A_1A_1$ ($H_1$)  &  $gg$, VBF, $VH$  &   \textcolor{green}{\checkmark}  300/fb &  $m_{A_1}<63$  \\
  $H_1\to A_1A_1$ ($H_2$)  &  $gg$, VBF, $VH$  & \textcolor{green}{\checkmark}  30/fb & $m_{A_1}<60$   \\
  $H_1\to A_1Z$  &   ---  &   \textcolor{red}{x} & \\
  $H_2\to A_1A_1$ ($H_1$)  &  $gg$, VBF  &  \textcolor{green}{\checkmark}  300/fb  &  $60< m_{A_1} <80$\\
  $H_2\to A_1A_1$ ($H_2$)  &  $gg$, VBF, $VH$  &   \textcolor{green}{\checkmark}  30/fb  &   $m_{A_1}<63$  \\
  $H_2\to A_1Z$   & ---  &   \textcolor{red}{x}  &  \\
  $H_3\to A_1A_1$  &  ---  &   \textcolor{red}{x}  & \\
  $H_3\to A_1Z$   & $gg$  &   \textcolor{green}{\checkmark}  300/fb  &   $60 < m_{A_1} <120$ \\\hline
  \end{tabular}
\caption{List of the $A_1$ production channels included in this
  study. The second column shows the production mechanisms of interest for the initial scalar, while the third column shows the
integrated luminosity at which the $A_1$ can be accessible at the LHC
in at least one of these combinations. In the fourth column we provide
the mass range within which a signature of $A_1$ can be established
in the given channel.}\label{tab:Channels}
\end{center}
\end{table}%

\section{Conclusions}
With its neat solution to the $\mu$ problem and improved ability to give a heavy enough Higgs boson, as compared to the MSSM, the NMSSM is a compelling model for new physics. The possible existence of light singlet states also holds great promise for new phenomenology at the LHC.

In this proceedings we have discussed work demonstrating that a light pseudoscalar can be discovered through $H_{1,2}\to A_1A_1$ in large parts of parameter space. This is especially interesting for the non-SM like of $H_1$ and $H_2$ as this might well be our only handle on such a scalar, i.e.\ seeing $H_{\rm non-SM}\to A_1A_1$ might not only be a way of discovering the pseudoscalar but also the scalar.

For pseudoscalars above the kinematic threshold of the above channels, our best chance might be $H_3\to A_1Z$, which is an interesting but poorly studied channel.

\end{document}